\documentclass[twoside,11pt]{article}

\usepackage{jmlr2e,amsmath}
\usepackage{breakurl}

\ShortHeadings{A Comparison of SVM Training GPU Implementations}{Van\v{e}k, Mich\'{a}lek, and Psutka}
\firstpageno{1}

\begin{document}

\title{A Comparison of Support Vector Machines Training GPU-Accelerated Open Source Implementations}

\author{\name Jan Van\v{e}k \email vanekyj@ntis.zcu.cz \\
       \addr New Technologies of the Information Society\\
       University of West Bohemia in Pilsen\\
       Technick\'{a} 8, 306 14 Plze\v{n}, Czech Republic
       \AND
       \name Josef Mich\'{a}lek \email orcus@kky.zcu.cz \\
       \addr Department of Cybernetics,
			 University of West Bohemia in Pilsen\\
       Technick\'{a} 8, 306 14 Plze\v{n}, Czech Republic
       \AND
       \name Josef Psutka \email psutka@ntis.zcu.cz \\
       \addr New Technologies of the Information Society,
			 University of West Bohemia in Pilsen\\
       Technick\'{a} 8, 306 14 Plze\v{n}, Czech Republic
			}

%\editor{}

\maketitle

\begin{abstract}%   <- trailing '%' for backward compatibility of .sty file
Last several years, GPUs are used to accelerate computations in many computer science domains. We focused on GPU accelerated Support Vector Machines (SVM) training with non-linear kernel functions. We had searched for all available GPU accelerated C++ open-source implementations and created an open-source C++ benchmark project. We modifed all the implementations to run on actual hardware and software and in both Windows and Linux operating systems. The benchmark project offers making a fair and direct comparison of the individual implementations under the same conditions, datasets, and hardware. In addition, we selected the most popular datasets in the community and tested them. Finally, based on the evaluation, we recommended the best-performing implementations for dense and sparse datasets.
\end{abstract}


%%%%%%%%%%%%%%%%%%%%%%%%%%%%%%%%%%%%%%%%%%%%%%%%%%%%%%%%%%%%%%%%%%%%%%%%%%%%%%%%
\section{Introduction}
%%%%%%%%%%%%%%%%%%%%%%%%%%%%%%%%%%%%%%%%%%%%%%%%%%%%%%%%%%%%%%%%%%%%%%%%%%%%%%%%

Training an SVM amounts to solving a quadratic programming problem. Very efficient solutions were developed for linear or linearized SVMs. However, nonlinear SVMs solvers are more computationally intensive. Therefore, GPUs computation power has been utilized widely. In \cite{Catanzaro2008} and \cite{Carpenter2009} nice speed-ups on dense data sets were reported with the comparison to LibSVM. However, a significant part of the speed-ups comes from a lack of the LibSVM performance that is not optimized in a high performance manner. A GPU multi-class SVM training was introduced in \cite{Herrero-Lopez2010}. In \cite{Li2011}, an SVM package under the same name like Catanzaro: GPUSVM was presented. It offers multi-class and cross-validation abilities. The most recent dense GPU implementation is WUSVM published in \cite{Tyree2014}. We also found three published implementations for sparse data. The regularized ELLPACK sparse matrix format was used in \cite{Lin2010}. In \cite{Cotter2011}, a data clustering was used to improve efficiency by the regularized sparse matrix. In \cite{Sopyla2012} the standard CSR format was used. All the mentioned GPU implementations are developed in nVidia CUDA. However, also other platforms were in focus: e.g. in \cite{Cadambi2009} FPGA was used, Intel Xeon Phi coprocessor was used in \cite{You2014}.

We discussed and compared the above mentioned GPU implementations in detail bellow. We developed a benchmark framework that ensures a fair comparison of the implementations. The benchmark offers data input/output and time measurement common for all the implementations. Therefore, all the implementations are compared with identical hardware, drivers, compiler, and only the GPU SVM training core varies. The benchmark project is open source and it is available at \url{https://github.com/OrcusCZ/SVMbenchmark}. The benchmark contains all compared implementations except KMLib, due to it being written in C\#. We also compared often out-dated projects on the same data, setup, and actual GPU hardware and results are in Section \ref{results}. 

\section{Included Open Source GPU Implementations}
\label{review}

\begin{itemize}
\item \textbf{LibSVM} published in \cite{Fan2005}. The LibSVM is a popular CPU-only implementation that we use as a reference. It is available at \url{https://www.csie.ntu.edu.tw/~cjlin/libsvm}. The LibSVM stores data in a sparse format. However, the dense variant is available also.
\item \textbf{GPU-LibSVM} published in \cite{Athanasopoulos2011} and available at \url{http://mklab.iti.gr/project/GPU-LIBSVM}. It is a modification of the dense variant of the LibSVM, where only the computation of the kernel matrix elements in only cross-validation mode is ported to GPU.
\item \textbf{GPUSVM} published in \cite{Catanzaro2008} and available at \url{https://code.google.com/p/gpusvm} is a more advanced CUDA implementation of a sequential minimal optimization (SMO).
\item \textbf{cuSVM} published in \cite{Carpenter2009} available at \url{http://patternsonascreen.net/cuSVM.html} is practically just a CUDA reimplementation of the LibSVM algorithm.
\item \textbf{MultiSVM} published in \cite{Herrero-Lopez2010} available at \url{https://code.google.com/p/multisvm} is the first GPU SVM implementation that allows a multi-class classification in the one-vs-all manner besides the two-class problems. A cross-task kernel caching technique is used to significantly reduce total amount of computations needed.
\item \textbf{gtSVM} published in \cite{Cotter2011} available at \url{http://ttic.uchicago.edu/~cotter/projects/gtsvm} uses sparse data format. It does not use SMO but it uses a larger working set of size 16. A clustering algorithm is used to regularize sparsity patterns in data and permits better memory access. The size of the clusters can be selected from two options: large or small that means 256 or 16 samples, respectively. In the results Section \ref{results} we marked the large clusters and the small clusters variants as "`gtSVM LC"' and "`gtSVM SC"', respectively.
\item \textbf{WUSVM} published in \cite{Tyree2014} is available at \url{https://github.com/wusvm}. A sparse primal SVM variant of the training algorithm is implemented in WUSVM. However the algorithm contains random shuffling of the training data by default that brings a stochastic component that produces models with variable performance in variable training times.
\item \textbf{KMLib} published in \cite{sopyla2015gpu} is written in C\# and uses CUDA.NET library. The libary supports several SVM kernels and also several sparse data formats, most notably Sliced EllR-T format introduced by this library.
\end{itemize}

\section{Results}
\label{results}

We tried to test all the implementations on the same datasets that were frequently used in the referenced publications. We used both dense and sparse datasets and coverted some small sparse datasets to dense form. We performed the two-class SVM training with RBF kernel function because it is supported by all the tested implementations. 
A complete list of used datasets and the training setup is in Table \ref{tab:datasets}. \textit{Epsilon} and \textit{Alpha} datasets come from the Pascal Large Scale Learning Challenge \url{http://largescale.ml.tu-berlin.de}. They are very large dense datasets and most implementations cannot handle them so we used only subsets of them. Scaling is used for \textit{Epsilon} and \textit{Alpha} datasets. The rest of the datasets come from the LibSVM data page: \url{https://www.csie.ntu.edu.tw/~cjlin/libsvmtools/datasets/}. \textit{Timit} and \textit{MNIST} are multi-class tasks, we converted them to binary task by classifying even-vs-odd class index. \textit{Adult} and \textit{Web} sets are the most popular SVM datasets and we used the biggest variants of the sets: \textit{Adult~a9a} and \textit{Web~w8a}. \textit{Cov1~Forest} is a set of cartographic variables for detection of the cover type. It is multi-class task but we used it as a detection of cover type~1 that is forest. \textit{20 Newsgroups}, \textit{RCV1}, and \textit{Real-Sim} are large sparse sets for the text categorization and the training data were used also as a testing ones for these three sets. We used a desktop PC with Intel Core i7-4790K, 4-core CPU clocked at 4.0GHz, 32 GB RAM and Pascal-based NVIDIA GTX 1080 with 2560~cores clocked at 1607~MHz and 8~GB GDDR5 with bandwidth 320~GB/s.

Dense datasets match better to the GPU architecture and speed-up over LibSVM may be very high because LibSVM processes data in sparse format. The results are in table \ref{tab:dense2}. The best results are bold. The reference LibSVM and LibSVM-dense implementations were set to use the maximum cache size equal to size of the GPU memory, so 8~GB was set in table \ref{tab:dense2}. The fastest implementation is the GPUSVM for \textit{Epsilon 40k} and \textit{Alpha 10k} and gtSVM for other datasets except \textit{Cov1~Forest}. cuSVM was able to outperform other implementations on \textit{Cov0~Forest} dataset. MultiSVM and gtSVM trained bad model with very low accuracy for \textit{Cov1~Forest}. wuSVM is inconsistent and each training takes different time. The table contains training time average observed in 10 tests. All other trained models gave the same accuracy on test sets as the LibSVM reference. 
Sparse datasets are harder to handle on GPUs. However, very high speed-ups were achieved for sparse sets also. The results are in table \ref{tab:sparse2}. The gtSVM implementation has two variants: the large clusters and the small clusters marked as "`gtSVM LC"' and "`gtSVM SC"', respectively. The large clusters variant gave better elapsed times for small sets with medium sparsity while the small cluster variant excelled on large and very sparse datasets. However, the \textit{Cov1~Forest} model gave very low accuracy. KMLib's training times were worse than gtSVM for all datasets except \textit{20 Newsgroups} and \textit{RCV1}, but it was able to train good model for \textit{Cov1~Forest}. The medium sparse datasets can be trained in the dense form also, however, the dense GPUSVM achieved slightly worse training times than the sparse gtSVM.

%%%%%%%%%%%%%%%%%%%%%%%%%%%%%%%%%%%%%%%%%%%%%%%%%%%%%%%%%%%%%%%%%%%%%%%%%%%%%%%%

%\acks{This work was supported by the Grant Agency of the Czech Republic, project No. GA\v{C}R GBP103/12/G084. }

%%%%%%%%%%%%%%%%%%%%%%%%%%%%%%%%%%%%%%%%%%%%%%%%%%%%%%%%%%%%%%%%%%%%%%%%%%%%%%%%

\begin{table*}[h]
\centering
\caption{List of used datasets with the main features and the training setup.}
\label{tab:datasets}
\makebox[\textwidth][c]{
%\begin{tabular}{|c||c|c|c|c|c|c|}
\begin{tabular}{ccccccc}
\hline\noalign{\smallskip}
Dataset & \# Training Samples & \# Test Samples & \# Dimensions & Dense / Sparse & C & Gamma \\
\noalign{\smallskip}
\hline
\noalign{\smallskip}
Epsilon 40k		&	40,000 	& 10,000 & 2,000 & Dense &  32 & 0.0001 \\
Alpha 10k 		& 10,000  & 50,000 &   500 & Dense & 512 & 0.002 \\
Timit					& 63,881  & 22,257 &    39 & Dense &  1 & 0.025 \\
Adult	a9a			& 32,561  & 16,281 &   123 & Sparse & 4 & 0.5 \\
Web w8a				& 49,749  & 14,951 &   300 & Sparse & 4 & 0.5 \\
MNIST					& 60,000  & 10,000 &   784 & Sparse & 1 & 0.02 \\
Cov1 Forest	  & 522,911  & 58,101 &   54 & Sparse & 3 & 1.0 \\
20 Newsgroups	& 19,996  & 19,996 & 1,335,191 & Sparse & 4 & 0.5 \\
RCV1		      & 20,242  & 677,399 & 47,236 & Sparse & 4 & 0.5 \\
Real-Sim		  & 72,309  & 72,309 & 20,958 & Sparse & 4 & 0.5 \\
\hline
\end{tabular}
}
\end{table*}

\begin{table*}[th]
\centering
\caption{Elapsed time of the SVM training in seconds on Pascal-based NVIDIA 1080 GPU and the dense datasets. WM means "`wrong model"}
\label{tab:dense2}
\makebox[\textwidth][c]{
%\begin{tabular}{|c||c|c|c|c|c|c|c|c|}
\begin{tabular}{ccccccccc}
\hline\noalign{\smallskip}
Dataset & Epsilon 40k & Alpha 10k & Timit & Adult a9a & Web w8a & MNIST & Cov1 Forest \\
\noalign{\smallskip}
\hline
\noalign{\smallskip}
%LibSVM		    &	4,646 	& 4,701  &  2,969 &  475  & 562  &  1,039 &  12,050 \\
LibSVM             &  1,526  &  57.5  &  159.1  &  69.3  &  237.2  &  275.6  &  6,990  \\
%LibSVM-dense  & 2,414 & 472  &  2,093  &  530  &  1,207 &  843 &  15,961 \\
LibSVM-dense  &  1,091  &  46.8  &  151.1  &  136.7  &  815.6  &  289.4  &  14,004  \\
%GPUSVM 		    &	\textbf{25.42} 	& \textbf{21.24}  &  \textbf{2.41} & \textbf{2.86}   &  \textbf{7.77} & \textbf{6.14}  & 273.4 \\
%cuSVM					& 30.46 &  54.91 & 12.80  & 11.02  & 16.66  & 9.89 & \textbf{233.3} \\
%cuSVM					& 28.88 &  56.01 & 9.43  & 10.33  & 22.37  & 8.82 & 232.6 \\
%MultiSVM			&	34.71	& 44.55  &  9.75  &  9.98  & 14.06  & 8.86  & \textit{149.6} WM  \\
%wuSVM         & 526.5 -- 607.1 & 129.8 -- 132.2 & 274.2 -- 382.8 & 267.1 -- 297.7 & 3.8 -- 4.1 & 118.8 -- 124.4 & ERR \\
    cuSVM & 39.25 & 101.38 & 17.04 & 17.45 & 22.02 & 11.92 & \textbf{265.27} \\
    gpuSVM & \textbf{28.04} & \textbf{26.26} & 2.86 & 3.36 & 7.58 & 5.89 & 301.04 \\
    multiSVM & 51.34 & 90.04 & 17.73 & 18.69 & 22.52 & 12.22 & WM\footnotemark[1] (191.05) \\
    wuSVM & 122.08 & 61.96 & 203.86 & 131.56 & 13.61 & 79.21 & 596.86 \\
    gtSVM LC & 47.80 & 31.45 & \textbf{2.43} & \textbf{3.19} & \textbf{4.02} & \textbf{3.72} & WM\footnotemark[1] (61.25) \\
    gtSVM SC & 77.22 & 36.38 & 3.17 & 3.37 & 4.43 & 4.55 & WM\footnotemark[1] (112.12) \\
\hline
\end{tabular}
}
\end{table*}

\begin{table*}[h]
\centering
\caption{Elapsed time of the SVM training in seconds on Pascal-based NVIDIA 1080 GPU and the sparse datasets. WM means "wrong model"}
\label{tab:sparse2}
\makebox[\textwidth][c]{
%\begin{tabular}{|c||c|c|c|c|c|c|c|}
\begin{tabular}{cccccccc}
\hline\noalign{\smallskip}
Dataset & Adult a9a & Web w8a & MNIST & Cov1 Forest & 20 Newsgroups & RCV1 & Real-Sim \\
\noalign{\smallskip}
\hline
\noalign{\smallskip}
%LibSVM		    &	 475	& 562  & 1,039 & 12,050  &  1,694 & 357 & 1,517   \\
LibSVM                         &   69.3  &  237.2  &  275.6  &  6,990  &  604.3  &  63.2  &  564.4  \\
%gtSVM LC			&	3.52 	&  \textbf{5.00} &  5.83 &  \textit{136.21} WM  & 2,649  & \textbf{20.86} &  \textbf{32.02}  \\
%gtSVM SC			&	\textbf{3.23} 	&  5.22 &  \textbf{4.86} &  \textit{84.14} WM  & 2,693 & 29.18 &  82.20  \\
    gtSVM LC & \textbf{3.30} & \textbf{4.03} & \textbf{3.69} & WM\footnotemark[1] (61.17) & 605.53 & 15.83 & 53.66 \\
    gtSVM SC & 3.44 & 4.27 & 4.60 & WM\footnotemark[1] (111.58) & 486.83 & 10.11 & \textbf{20.81} \\
    KMLib & 18.65 & 19.30 & 21.55 & \textbf{874.65} & \textbf{145.01} & \textbf{10.02} & 55.80 \\
\hline
\end{tabular}
}
\end{table*}

%%%%%%%%%%%%%%%%%%%%%%%%%%%%%%%%%%%%%%%%%%%%%%%%%%%%%%%%%%%%%%%%%%%%%%%%%%%%%%%%

%%%%%%%%%%%%%%%%%%%%%%%%%%%%%%%%%%%%%%%%%%%%%%%%%%%%%%%%%%%%%%%%%%%%%%%%%%%%%%%%

\newpage
\vskip 0.2in
\bibliography{main}

\begin{thebibliography}{13}
\providecommand{\natexlab}[1]{#1}
\providecommand{\url}[1]{\texttt{#1}}
\expandafter\ifx\csname urlstyle\endcsname\relax
  \providecommand{\doi}[1]{doi: #1}\else
  \providecommand{\doi}{doi: \begingroup \urlstyle{rm}\Url}\fi

\bibitem[Athanasopoulos and Dimou(2011)]{Athanasopoulos2011}
A~Athanasopoulos and A~Dimou.
\newblock {GPU acceleration for support vector machines}.
\newblock \emph{WIAMIS 2011}, 2011.

\bibitem[Cadambi et~al.(2009)Cadambi, Durdanovic, Jakkula, Sankaradass,
  Cosatto, Chakradhar, and Graf]{Cadambi2009}
Srihari Cadambi, Igor Durdanovic, Venkata Jakkula, Murugan Sankaradass, Eric
  Cosatto, Srimat Chakradhar, and Hans~Peter Graf.
\newblock {A Massively Parallel FPGA-Based Coprocessor for Support Vector
  Machines}.
\newblock \emph{2009 17th IEEE Symposium on Field Programmable Custom Computing
  Machines}, pages 115--122, 2009.
\newblock \doi{10.1109/FCCM.2009.34}.

\bibitem[Carpenter(2009)]{Carpenter2009}
Austin Carpenter.
\newblock {cuSVM: A CUDA implementation of support vector classification and
  regression}.
\newblock \emph{patternsonscreen. net/cuSVMDesc. pdf}, pages 1--9, 2009.

\bibitem[Catanzaro et~al.(2008)Catanzaro, Sundaram, and Keutzer]{Catanzaro2008}
Bryan Catanzaro, N~Sundaram, and Kurt Keutzer.
\newblock {Fast support vector machine training and classification on graphics
  processors}.
\newblock \emph{Proceedings of the 25th International Conference on Machine
  Learning}, pages 104--111, 2008.

\bibitem[Cotter et~al.(2011)Cotter, Srebro, and Keshet]{Cotter2011}
Andrew Cotter, N~Srebro, and J~Keshet.
\newblock {A GPU-tailored approach for training kernelized SVMs}.
\newblock \emph{Proceedings of the 17th ACM SIGKDD international conference on
  Knowledge discovery and data mining}, pages 805--813, 2011.

\bibitem[Fan et~al.(2005)Fan, Chen, and Lin]{Fan2005}
RE~Fan, PH~Chen, and CJ~Lin.
\newblock {Working set selection using second order information for training
  support vector machines}.
\newblock \emph{The Journal of Machine Learning Research}, 6:\penalty0
  1889--1918, 2005.

\bibitem[Herrero-Lopez et~al.(2010)Herrero-Lopez, Williams, and
  Sanchez]{Herrero-Lopez2010}
S~Herrero-Lopez, JR~Williams, and Abel Sanchez.
\newblock {Parallel multiclass classification using SVMs on GPUs}.
\newblock \emph{Proceedings of the 3rd Workshop on General-Purpose Computation
  on Graphics Processing Units}, pages 2--11, 2010.

\bibitem[Li et~al.(2011)Li, Salman, and Test]{Li2011}
Qi~Li, Raied Salman, and Erik Test.
\newblock {GPUSVM: a comprehensive CUDA based support vector machine package}.
\newblock \emph{Open Computer Science}, pages 1--22, 2011.

\bibitem[Lin and Chien(2010)]{Lin2010}
Tsung-Kai Lin and Shao-Yi Chien.
\newblock {Support Vector Machines on GPU with Sparse Matrix Format}.
\newblock In \emph{2010 Ninth International Conference on Machine Learning and
  Applications}, pages 313--318. Ieee, December 2010.
\newblock ISBN 978-1-4244-9211-4.
\newblock \doi{10.1109/ICMLA.2010.53}.

\bibitem[Sopyla et~al.(2012)Sopyla, Drozda, and G\'{o}recki]{Sopyla2012}
K~Sopyla, Pawel Drozda, and P~G\'{o}recki.
\newblock {SVM with CUDA Accelerated Kernels for Big Sparse Problems}.
\newblock \emph{Artificial Intelligence and Soft Computing}, pages 439--447,
  2012.

\bibitem[Sopyla and Drozda(2015)]{sopyla2015gpu}
Krzysztof Sopyla and Pawel Drozda.
\newblock {GPU Accelerated SVM with Sparse Sliced EllR-T Matrix Format}.
\newblock \emph{International Journal on Artificial Intelligence Tools},
  24\penalty0 (01):\penalty0 1450012, 2015.

\bibitem[Tyree et~al.(2014)Tyree, Gardner, Weinberger, Agrawal, and
  Tran]{Tyree2014}
Stephen Tyree, Jacob~R. Gardner, Kilian~Q. Weinberger, Kunal Agrawal, and John
  Tran.
\newblock Parallel support vector machines in practice.
\newblock \emph{arXiv preprint arXiv:1404.1066}, 2014.

\bibitem[You et~al.(2014)You, Song, Fu, Marquez, Dehnavi, Barker, Cameron,
  Randles, and Yang]{You2014}
Yang You, Shuaiwen~Leon Song, Haohuan Fu, Andres Marquez, Maryam~Mehri Dehnavi,
  Kevin Barker, Kirk~W. Cameron, Amanda~Peters Randles, and Guangwen Yang.
\newblock {MIC-SVM: Designing a Highly Efficient Support Vector Machine for
  Advanced Modern Multi-core and Many-Core Architectures}.
\newblock \emph{2014 IEEE 28th International Parallel and Distributed
  Processing Symposium}, pages 809--818, May 2014.
\newblock \doi{10.1109/IPDPS.2014.88}.

\end{thebibliography}

\end{document}